\documentclass[useAMS,usenatbib,flqen]{mn2e}
\usepackage{psfig,amsmath}
\usepackage[dvips]{graphicx}
\usepackage{amssymb}

\def\xmm{{\sl XMM-Newton}}

\def\xte{{\sl RXTE}}

\def\chandra{{\sl Chandra}}
\def\me{{ $\dot{m}_{E}$ }}

\title[Accretion state of NGC 3783]{Timing evidence in determining the accretion state of the Seyfert galaxy NGC 3783}  
\author[D. P. Summons et al]{D. P. Summons$^{1,3}$\thanks{E-mail: dps@astro.soton.ac.uk}, 
P. Ar\'{e}valo$^{1}$ , I. M. M$^{\mathrm{c}}$Hardy$^{1}$, P. Uttley$^{2}$ and A. Bhaskar$^{3}$ \\
$^1$School of Physics and Astronomy, University of Southampton, Southampton SO17 1BJ, UK \\ 
$^2$Astronomical Institute `Anton Pannekoek', University of Amsterdam, Kruislaan 403, 1098 SJ, Amsterdam, the Netherlands\\
$^3$School of Engineering Science, University of Southampton, Southampton SO17 1BJ, UK
}
\begin{document}
\date{Received /Accepted}
\pagerange{\pageref{firstpage}--\pageref{lastpage}} \pubyear{2006}

\maketitle
\label{firstpage}

\begin{abstract}
Previous observations with the \textit{Rossi X-ray Timing Explorer}
(\xte) have suggested that the power spectral density (PSD) of NGC
3783 flattens to a slope near zero at low frequencies, in a similar
manner to that of Galactic black hole X-ray binary systems (GBHs) in
the `hard' state. The low radio flux emitted by this object, however,
is inconsistent with a hard state interpretation. The accretion rate
of NGC 3783 ($\sim 7\%$ of the Eddington rate) is similar to that of
other AGN with `soft' state PSDs and higher than that at which the GBH
Cyg~X-1, with which AGN are often compared, changes between `hard' and
`soft' states ($\sim 2\%$ of the Eddington rate). If NGC 3783 really
does have a `hard' state PSD, it would be quite unusual and would indicate that AGN and GBHs are not quite as
similar as we currently believe.  Here we present an improved X-ray
PSD of NGC 3783, spanning from $\sim 10^{-8}$ to $\sim 10^{-3}$ Hz,
based on considerably extended (5.5 years) \xte\ observations combined
with two orbits of continuous observation by
\xmm. We show that this PSD is, in fact, well fitted by a `soft' state
model which has only one break, at high frequencies. Although a `hard'
state model can also fit the data, the improvement in fit by adding a
second break at low frequency is not significant. Thus NGC 3783 is not
unusual. These results leave Arakelian 564 as the only AGN which shows
a second break at low frequencies, although in that case the very high
accretion rate implies a `very high', rather than `hard' state
PSD. The break frequency found in NGC 3783 is consistent with the 
expectation based on comparisons with other AGN and GBHs, given its
black hole mass and accretion rate.
\end{abstract}


\section{INTRODUCTION}

Super-massive black holes in active galactic nuclei (AGN) and Galactic
stellar-mass black hole X-ray binary systems (GBHs) both display
aperiodic X-ray variability which may be quantified by calculating the
power spectral densities (PSDs) of the X-ray light curves. The PSDs
can typically be represented by red-noise type power laws
(i.e. $P(\nu)$, the power at frequency $\nu$, $\propto \nu^{\alpha}$
where $\alpha \sim$ -1) with a bend or break (to $\alpha \leq$ -2) at
a characteristic PSD frequency. The 
time-scale, corresponding to the bend-frequency, scales approximately linearly with black hole mass from AGN
to GBHs
\citep{McHardy:1988kx,Edelson:1999uq,Uttley:2002zw,
Uttley:2005tq,Markowitz:2003gm, McHardy:2004hv,McHardy:2005pe}, albeit
with some scatter.  However, the scatter is entirely accounted for by
variations in accretion rate, allowing scaling between AGN and GBHs on
time-scales from $\sim$ years to $\sim$ ms \citep{McHardy:2006fk}.

GBHs are observed in a number of distinct X-ray spectral states which
also have distinct X-ray timing properties.  Two common states are the
low/hard (hereafter `hard') and high/soft (hearafter `soft')
states. In the hard state, the energy-spectrum is dominated by a
highly variable power law component and the PSDs are well fitted by
multiple broad Lorentzians. For use in AGN, where signal/noise is
lower than in GBHs, this PSD shape can be approximated by a
double-bend power law with slopes $\alpha=0$, -1  and -2, from low to
high frequency, where the high- and low-frequency bends correspond to
the strongest peaks in the Lorentzian parameterisation. The breaks are
typically separated by only one to two decades in frequency.  In the
soft state, the energy spectrum is dominated by an approximately
constant thermal disc component which extends into the X-ray band in
GBHs but which in AGN is shifted down to the optical/UV
band. Therefore, a meaningful comparison between the PSDs of soft
state GBH and AGN can only be made in cases where the GBH power-law
emission is strong enough to show significant variability.  Such GBHs
are rare and the best example is Cyg~X-1 which shows a `1/f' PSD over
many decades of frequencies \citep{Reig:2002ys}.  The soft state is
distinguished from the hard state by having only one, high frequency,
break in this power law, from slope -1 to -2.

It has been suggested that this pure simple broken or cut-off
power-law PSD shape is unique to the soft state of Cyg~X-1, which is a
persistent source. However in transient GBHs with similar X-ray
spectra, the power law PSD component may be seen in combination with
broad Lorentzian features \citep{Done:2005fk}.  \cite{Axelsson:2006fk} also note that a mixed power law plus Lorentzian PSD is
also present in Cyg~X-1 in lower luminosity, harder spectral states,
but as the luminosity rises the Lorentzian features weaken and the
power law PSD component strengthens until, in the softest state, it
completely dominates.  Since the softest
spectral states of transient GBHs are dominated by
constant disc emission we cannot determine whether they show a similar PSD shape
to Cyg~X-1.

However, a direct comparison of transient GBHs and
Cyg~X-1 is complex, since the transients show much larger luminosity
changes, and complex hysteresis effects in spectral hardness versus
luminosity (e.g. \citealt{Homan:2001fk}, \citealt{Belloni:2005uq}) which are not
seen in Cyg~X-1.  Therefore it is not clear that one can compare
timing properties between Cyg~X-1 and transient GBHs simply as a
function of observed X-ray spectrum.

None the less, it is still interesting that the X-ray spectrum of Cyg~X-1
never becomes totally disc-dominated, and always contains a relatively
strong variable component whose PSD resembles that of X-ray bright AGN. If
variability originates, at least partly, in the disc, so power spectral
shape is related to the disc structure, that structure might be severely
disrupted during outbursts, thereby suggesting a possible difference
between the persistent Cyg~X-1 and the transient GBH sources. The
similarities between the PSDs of Cyg~X-1 and AGN may also be related to
the possible similarities in accretion flows between AGN and Cyg~X-1 noted
by \cite{Done:2005fk}.

To date, NGC~3783 and the Narrow Line Seyfert~1 Galaxy (NLS1) Ark~564
are the only AGN with suggested second, low-frequency breaks in their
PSDs (i.e. similar to low/hard GBHs) and are both commonly referred to as being unusual (e.g.
\citealt{Done:2005fk}). The power spectral evidence for a second break in
the case of Ark~564 is very strong
(\citealt{Pounds:2001nx,Papadakis:2002fk,Markowitz:2003gm}, M$^{\rm
c}$Hardy et al. in prep.). Of all the AGN with good timing data,
Ark~564 shows the highest accretion-rate (possibly super-Eddington) so
it would not be surprising if it were in an unusual state, e.g. the
`very high' state where the PSD, in GBHs, also displays two distinct
breaks.  The properties of NGC~3783, on the other hand, are similar to
those of AGN with proven soft-state PSDs (e.g. NGC~3227, NGC~4051
\citealt{McHardy:2004hv}, MCG-6-30-15
\citealt{McHardy:2005pe}), and in particular it is radio quiet (e.g. \citealt{Reynolds:1997vn}).
In the hard state, GBHs are strong radio sources whereas in the soft
state the radio emission is quenched \citep{Corbel:2000zr,Fender:2001ys,Kording:2006ly}. We
also note that NGC~3783 has a more moderate accretion rate than Ark~564 ($\sim 7
\%$), and more similar to the other AGN mentioned above, and
Cyg~X-1 changes from the hard to the soft state at around
2\% of the Eddington accretion rate (i.e. \me =0.02)
(\citealt{Pottschmidt:2003kx,Wilms:2006uq,Axelsson:2006fk}). These two
facts do not lie easily with a hard state identification of NGC~3783.
Thus it would be surprising, and might indicate that our current ideas
regarding the scaling between AGN and GBHs are not entirely correct,
if NGC~3783 were proven to have a hard state PSD. It is therefore
important to determine whether NGC~3783 does have a second, low
frequency, break in its PSD or not.

\cite{Markowitz:2003gm} recognised the presence of a break in 
the $2-10$ keV PSD of NGC~3783 at $4\times 10^{-6}$ Hz and found
provisional evidence for a second lower-frequency break at $\sim$
2$\times 10^{-7}$ Hz. Specifically, \cite{Markowitz:2003gm} rejected
the possibility that the PSD is described by a single-break power law
with low-frequency slope -1, similar to other AGN, at the 98\%
confidence level.  In this paper we re-investigate the evidence for
the second break in the PSD of NGC~3783, using new long-term
monitoring data that covers the frequency range where the break
appears to be.  By including additional \xte\ archival data spanning
several years, along with short time-scale observations by \xmm, we
will demonstrate that the improved PSD is perfectly compatible with a
single-bend power law, consistent with the behaviour of the other
moderately-accreting Seyferts.  In Section 2 we describe the
observations and the methods by which we extract the \xte\ and \xmm\
light curves.  In Sections 3 we discuss the PSD of NGC 3783 as
produced from the \xte\ and \xmm\ observations, and compare it with
various PSD models.  In Section 4 we briefly review the implications
of our observations.

\section{OBSERVATIONS AND DATA REDUCTION}

\subsection{\xte\ Data Reduction}

From 1999 to 2006, NGC~3783 has been the target of various monitoring
campaigns with \xte. These campaigns have consisted of short, $\sim$1 ks
duration, observations with the proportional counter array (PCA,
\citealt{Zhang:1993ly}). We have analysed the archival
PCA STANDARD-2 data and our own proprietary data with FTOOLS v6.0.2
using standard extraction methods. We use data from the top layers of
PCUs 0 and 2 up to 2000 May 12 and only top layer PCU 2 data from
observations after this date.  The remaining PCUs were not used due to
repeated breakdowns.

Data were selected according to the standard `goodtime' criteria, i.e.
target elevation $<10^{\circ}$, offset pointing $<0.02^{\circ}$, and
electron contamination $<0.1$. The background was simulated with the
L7 model for faint sources using PCABACKEST v3.0.  The response
matrices for each PCA observation were calculated using PCARSP v10.1.
The final $2-10$ keV fluxes were calculated using XPSEC v12.2.1 by fitting a
power law with galactic absorption to the PHA data.

\begin{figure}
\psfig{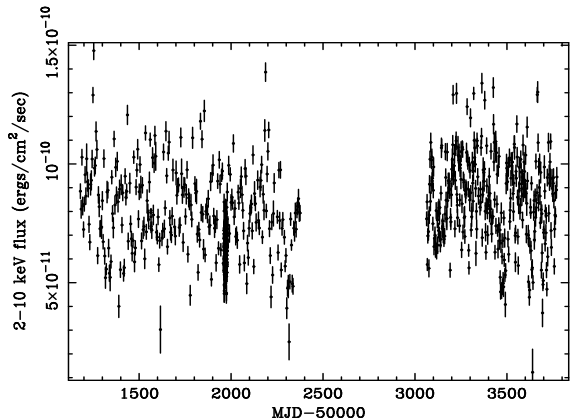}
\caption{\xte\ long-term light curve of NGC~3783 in the 2-10keV band.}
\label{longlc}
\end{figure}

The \xte\ data used in our analysis, together with the sampling
patterns, are listed in Table \ref{sample} and displayed in Fig.
\ref{longlc}.  The early data (to MJD 52375) with 4~d sampling,
together with the 20~d period of 3~h sampling already presented by
\cite{Markowitz:2003gm} are followed, after a 2 year gap, by our
new long-term monitoring, with 2~d sampling. As the gap is large
compared to the duration of each monitoring campaign we will include
data from each monitoring campaign as separate lightcurves in our fits.

\begin{table*}
\centerline{}
\begin{tabular}{|r||c||c||r|} \hline
{\bf Light curve } & {\bf Sampling interval} & {\bf Observation length} & {\bf Date Range [MJD]} \\ 
\hline
\xte\ Long-term 1 & $\sim$4.36 days & 1194.6 days & 51180.5--52375.1 \\
\xte\ Long-term 2 & $\sim$2.1 days & 928.3 days & 53063.4--53991.6 \\
\xte\ Intense monitoring & $\sim$3.2 hours & 19.9 days & 51960.1--51980.1 \\
\xmm\ observations (2 orbits) & 200-s & 3.2 days & 52260.8--52264.0 \\
\hline
\end{tabular}
\caption{Summary of the \xte\ and \xmm\ light curves used in the analysis of NGC~3783, including their sampling frequency and date range.}
\label{sample}
\end{table*}

\begin{figure}
\psfig{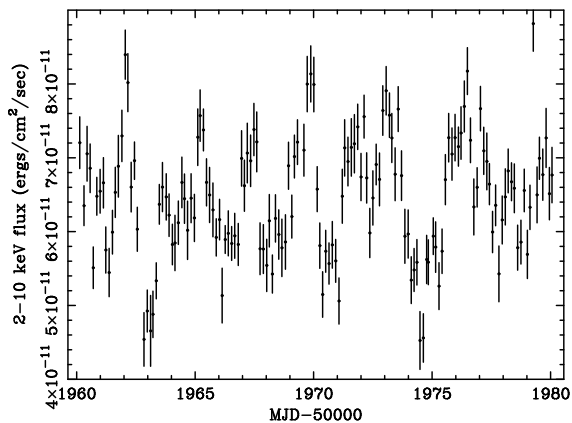}
\caption{\xte\ intense sampling light curve in the 2-10 keV band of NGC~3783.}
\label{intlc}
\end{figure}

\subsection{\xmm\ Data Reduction}

NGC~3783 was observed by \xmm\ during revolutions 371 and 372, between
2001 December 17 and 2001 December 21. Temporal analysis of these data
were first presented by \cite{Markowitz:2005zb} who discusses the
coherence, frequency-dependent phase lags, and variation of high
frequency PSD slope with energy. Here we use these data to constrain
the high frequency part of the overall long and short timescale PSD.
We used data from the European Photon Imaging Cameras (EPIC) PN and
MOS2 instruments, which were operated in imaging mode. MOS1 was
operated in Fast Uncompressed Mode and we do not use those data here.  The PN camera was operated in Small Window mode, using
the medium filter. Source photons were extracted from a circular
region of $40 \arcsec$ radius and the background was selected from a
source-free region of equal area on the same chip. We selected
single and double events, with quality flag=0.  The MOS2 camera was
operated in the Full Window mode, using the medium filter.  We
extracted source and background photons using the same procedure as
for the PN data and selected single, double, triple and quadruple
events. These data showed no serious pile-up when tested with the {\it
XMM-SAS} task {\it epatplot}.

\begin{figure}
\psfig{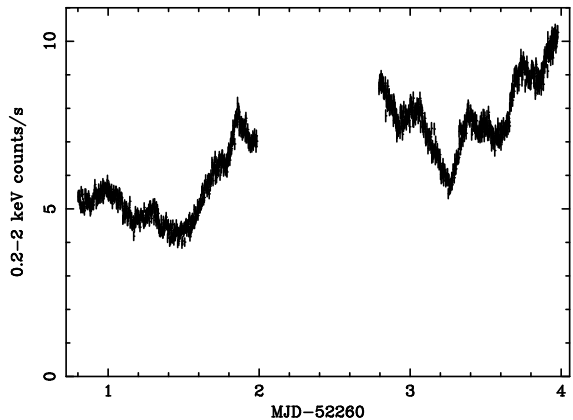}
\caption{\xmm\ light curve in the 0.2--2 keV band of NGC~3783.}
\label{xmmlcfig}
\end{figure}

We constructed light curves, for each detector and orbit, in the
0.2--2, 2--10 and 4--10 keV energy bands. We filled in the $\sim 5$ ks
gap in the middle of orbit 371 light curves, and some other much
smaller gaps, by interpolation and added Poisson noise. The resulting
PN and MOS2 continuous light curves were then combined to produce the
final light curves for each orbit. The combined, background
subtracted, average count rates in the 0.2--10 keV band were 11.8 c/s
for orbit 371 and 15.8 c/s for orbit 372, and the 0.2-2 keV combined
light curve is shown in Fig. \ref{xmmlcfig}. Poisson noise dominates
the PSD on timescales shorter than 1000s, so the light curves were
binned into 200s bins.

\section{Power spectral Modelling}

\subsection{Combining \xte\ and \xmm\ data} 

To determine the PSD over the largest possible
frequency range we combine the \xte\ and \xmm\ data. In GBHs the break-frequency and slope of the PSD below the break appear to be independent of the chosen energy band \citep{Cui:1997zr,Churazov:2001le,Nowak:1999ys,Revnivtsev:2000vn,McHardy:2004hv}.  On the other hand, the PSD normalisation and the slope above the break are often energy-dependent  \citep{Markowitz:2005zb}. Therefore, when combining data from
different instruments, it is preferable to use similar energy ranges.
The \xte\ data are in the 2--10 keV band and, for NGC~3783, that band
has a median photon energy of 5.7 keV. The \xmm\ band with the same
median photon energy is 4.1--10 keV. However the count rate in that
\xmm\ band is low (2 c/s) so we only detect significant source power
above the Poisson noise level at frequencies below $10^{-4}$Hz.  To
probe higher frequencies we can use the 0.2--2 keV \xmm\ data (8.8 c/s)
but we must re-scale its PSD normalisation to that of the 4--10 keV
PSD. We determined the scaling correction by producing PSDs in both
energy bands and fitting the same bending power law model to the
noise-subtracted data. On the assumption that the PSD shape below the
high frequency break is energy-independent, the combined \xte\ 2--10
keV and \xmm\ 0.2--2 keV PSD will then have the shape of the 0.2--2 keV
PSD.

\subsection{Monte Carlo simulations}
\label{MCsims}
We use the Monte Carlo technique of \cite{Uttley:2002zw} (PSRESP), to
estimate the underlying PSD parameters in the presence of sampling
biases.  In this method we first calculate the observed (or `dirty')
PSD, in parts, from the observed lightcurves, using the Discrete
Fourier Transform.  Here the PSD estimates are binned in bins of width
1.3$\nu$, where $\nu$ is the starting frequency by taking the average
of the log of power \citep{Papadakis:1993xl}. We require a minimum
of 4 PSD estimates per bin. We then compare
simultaneously the dirty PSDs from each lightcurve with various model
PSDs derived from lightcurves simulated with the same sampling pattern
as the real observations.  We alter the model
parameters to obtain the best fit for any given model.  We refer the
reader to \cite{Uttley:2002zw} for a full discussion of the method.

For each set of chosen underlying-PSD model parameters, we simulate
red-noise light curves, as described by \cite{Timmer:1995qj}.  The
\xte\ light curves are simulated with time resolutions of 10.5 h, 5.0 h
and 18 m for the first and second long time-scale and the medium
time-scale light curves respectively.  The simulated resolution, which
is 10 times shorter than the typical sampling intervals of the real
observations, given in column 2 of Table \ref{sample}, is to take into
account the effect of aliasing.  These simulated light curves were
resampled and binned to match the real NGC~3783 observations.  \xmm\
light curves were simulated with 200-s resolution, as at shorter
time-scales the underlying varability power is negligible compared to
the Poisson noise, so aliasing does not play a role. The Poisson noise
level was not subtracted from the observed PSD but was added to the
simulated PSDs. To reproduce the effect of red-noise leak, each light
curve was simulated to be $\sim$300 times longer than the real
observation, and was then split into sections, constituting 300
simulated light curves for each observed lightcurve. The simulated
model average PSD is evaluated from this ensemble of PSD realisations,
and the errors are assigned from the rms spread of the realisations
within a frequency bin.

\begin{figure}
\psfig{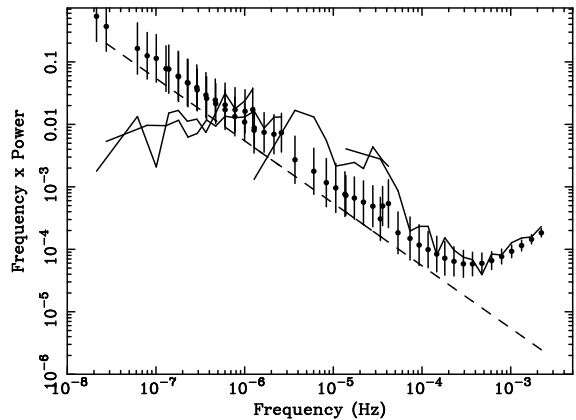}
\caption{The best fit unbroken power law PSD of the combined \xte\ and
\xmm\ data.  The solid lines represent the observed data and the points
with error bars exhibit the biased model and the spread
in individual realisations of the model.  The dashed line is the
underlying model used to generate the simulated PSD.  The three lowest
frequency data sets are from \xte\ observations and the high frequency
data set ($\sim$$10^{-5}$ Hz) is from the \xmm\ observations.  Note that the rise in power at the highest frequencies is due to the photon Poisson noise.}  
\label{novb} \end{figure}

We present the results of several PSD model fits in an attempt to
quantify the underlying model shape that best describes the PSD of
NGC~3783, and associate an acceptance probability with each model. We
initially test a simple unbroken power law model.  We next fit a power law
with a single-bend in the PSD, and then a model incorporating a
double-bend. We also fit a single-bend power law with a Lorentzian
component.

\subsection{Unbroken power law model}

To begin, we fitted a simple power law model to the data of the form:

\begin{equation*}
P(\nu)=A\left(\frac{\nu}{\nu_0}\right)^{\alpha}
\end{equation*}

where $A$ is the normalisation at a frequency $\nu_0$, and $\alpha$ is
the power law slope.  We made 900 simulations and in Fig. \ref{novb}
we show the best fit plotted in $\nu \times P_{\nu}$, which has a PSD slope of -$2.1$. However, the fit
is poor and this model can be rejected with a probability $>$ 99 \%,
or $\gtrsim$ 3 $\sigma$.

\begin{figure}
\psfig{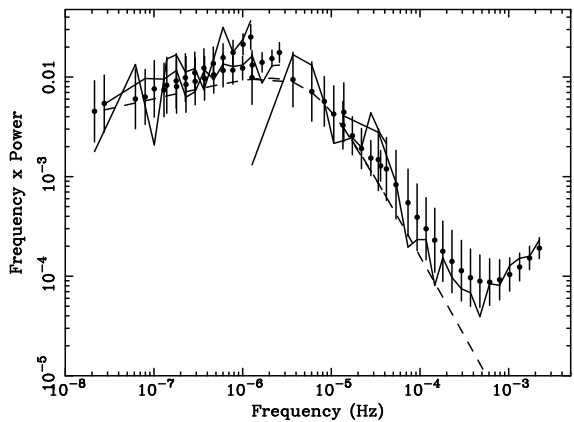}
\caption{The best fit single-bend power law PSD of the combined \xte\ and \xmm\ data.   The various lines represent the same data as seen in Fig. \ref{novb}. }
\label{sinpsd}
\end{figure}

\begin{figure}
\psfig{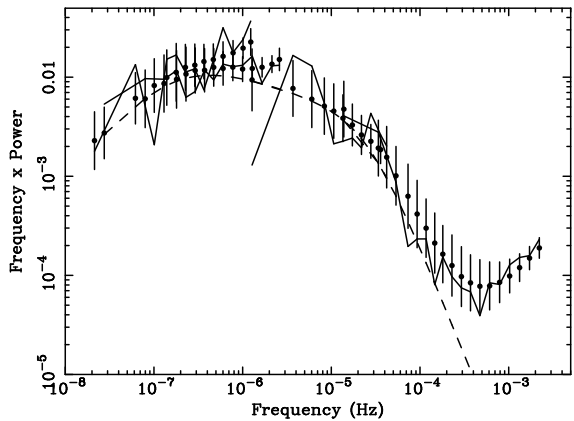}
\caption{The best fit double-bend power law PSD of the combined \xte\ and \xmm\ data.  The various lines represent the same data as seen in Fig. \ref{novb}.}
\label{doupsd}
\end{figure}

\subsection{Single-bend power law model}

Here we fit a single-bend power law to the data. This model
best describes the PSD of Cyg X-1 in the high/soft state, and provides
a good fit to the PSDs of the AGN NGC~4051 and MCG--6-30-15
\citep{McHardy:2004hv,McHardy:2005pe}.

\begin{equation*}
P(\nu)=\frac{A\:\nu^{\alpha_L}}{1+\left(\frac{\nu}{\nu_B}\right)^{\alpha_L-\alpha_H}}
\end{equation*}

\begin{figure}
\psfig{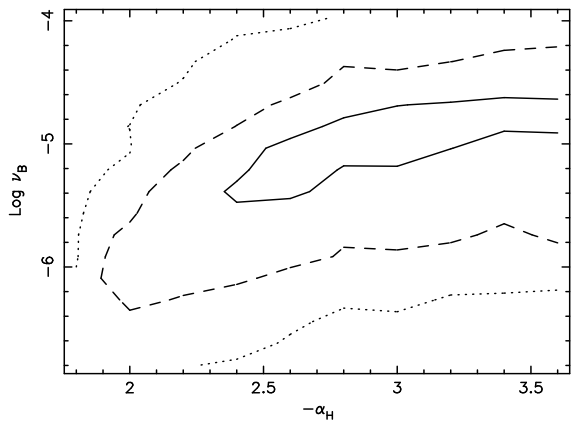}
\caption{Single-bend power law model: 68, 90, and 99\% confidence contours for the bend frequency, $\nu_B$, and the high frequency slope, $-\alpha_H$, for the single-bend power law fit to the combined \xte\ and \xmm\ PSD.}
\label{singlecon13}
\end{figure}

\begin{figure}
\psfig{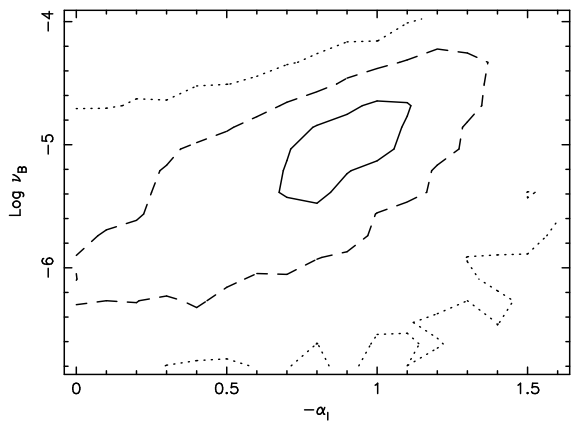}
\caption{Single-bend power law model: 68, 90, and 99\% confidence contours for the bend frequency, $\nu_B$, and the low frequency slope, $-\alpha_L$, for the single-bending power law fit to the combined \xte\ and \xmm\ PSD.}
\label{singlecon14}
\end{figure}

Fig. \ref{sinpsd} presents the observed PSD fitted with a
single-bend power law model, for which a good likelihood of acceptance
is obtained (P = 44 \%).  The best fit bend-frequency is
$\nu_B=6.2^{+40.6}_{-5.6}\times 10^{-6}$ Hz, the high-frequency slope
is $\alpha_H=-2.6^{+0.6}_{-*}$, and the low-frequency slope is
$\alpha_L=-0.8^{+*}_{-0.5}$.  The errors are  90\%
confidence limits, an asterisk indicates that the limit is unconstrained.    
For $\alpha_H$ the best fit value is well within the searched
parameter space but the degeneracy produced by red-noise leak in the
probability at high values of $\alpha_H$, means that the upper limit 
is not constrained at the 90\% confidence level.  
The confidence contours for the main interesting parameters are plotted in
Figs. \ref{singlecon13} and \ref{singlecon14}. 
Table \ref{results} shows the single-bend power law best fit parameters to the data.
The best fit single-bend frequency obtained here is
consistent with the value found by \cite{Markowitz:2003gm}

\begin{table*}
\centerline{}
\begin{tabular}{|c||c||c||c||c||c||c|c} \hline
{\bf Model} & {\bf Normalisation} & {\bf $\alpha_{H}$} & {\bf $\alpha_{I}$} &{\bf $\alpha_{L}$}& {\bf $\nu_H$} & {\bf $\nu_L$} & {\bf Acceptance}\\
{}& {($a$)} & {} & & {} & {(Hz)} & {(Hz)} & {(\%)}\\
\hline
Single-bend & $1.5\times 10^{-4}$ & $-2.6^{+0.6}_{-1.0}$ &NA& $-0.8^{+0.8}_{-0.5}$ &$6.2^{+40.6}_{-5.6}\times 10^{-6}$ & NA & 44.4\\
Double-bend & $1.0\times 10^{2}$ & $-3.2^{+1.2}_{-*}$ & $-1.3^{+*}_{-*}$ & 0.0&  $2.6^{+*}_{-*}\times 10^{-5}$ &$1.7^{+*}_{-*}\times 10^{-7}$ & 63.9\\
\hline
\end{tabular}
\caption{Best fit model parameters for the examined models to the combined \xte\ and \xmm\ PSD of NGC~3783.  The errors on the single- and double-bend fits are calculated from the 90\% confidence intervals.   The bend-frequency for the single-bend model, $\nu_B$, is denoted here as $\nu_H$.  An asterisk indicates that the limit is unconstrained.}
\label{results}
\end{table*}

\subsection{Double-bend power law model}

\cite{Markowitz:2003gm} provide tentative evidence that a second,
lower, frequency break exists in the PSD of NGC~3783.  Thus, we also
fitted a more complex double-bend power law model to see if the
goodness-of-fit is improved significantly. The double-bend power law
model is given by:

\begin{equation*}
P(\nu)=\frac{A\:\nu^{\alpha_L}}{\left[1+\left(\frac{\nu}{\nu_L}\right)^{\alpha_L-\alpha_I}\right]\left[1+\left(\frac{\nu}{\nu_H}\right)^{\alpha_I-\alpha_H}\right]},
\end{equation*}

where $\alpha_I$ is the intermediate-frequency slope and $\nu_L$ and
$\nu_H$ are the low and high bend-frequencies respectively. We fixed
the low-frequency slope to zero, to avoid making the simulations
computationally prohibitive, and because a low-frequency slope of zero would allow
the best qualitative comparison to the low state of Cyg X-1
\citep{Nowak:1999ys}.

Fig. \ref{doupsd} presents the same observed PSD as in
Fig. \ref{sinpsd}, but fitted with the double-bend power law model.  A
good likelihood of acceptance is obtained (P=64 \%).  The best-fitting
high bend-frequency is $\nu_H=2.6^{+*}_{-*}\times 10^{-5}$ Hz, the
high-frequency slope is $\alpha_H=-3.2^{+1.2}_{-*}$, the
intermediate-frequency slope is $\alpha_I=-1.3^{+*}_{-*}$, the
low-frequency bend is $\nu_L=1.7^{+*}_{-*}\times 10^{-7}$ Hz.  As
before, we use 90\% confidence limits.  The added parameters allow
extra freedom to find better fit probabilities for any given set of
double-bend parameters. For this reason, the contour levels cover
larger ranges in the parameter space and therefore, most of the 90\%
contours in our double-bend fit remain unbounded over the fitted
parameter space.  The high-frequency slope is subject to the same problems as in the
single-bend model.  Table
\ref{results} contains a summary of the best-fitting model parameters.
 
\begin{figure}
\psfig{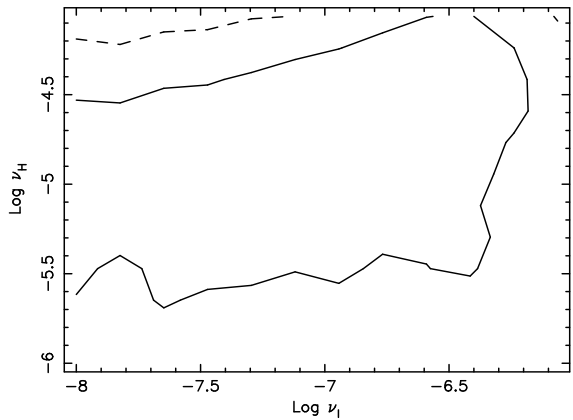}
\caption{Double-bend power law model: 68, and 90\% confidence contours for the high bend-frequency, $\nu_H$, and the low bend-frequency, $\nu_L$, for the double-bend power law fit to the combined \xte\ and \xmm\ PSD.}
\label{doublecon12}
\end{figure}

The best-fitting low-frequency bend is found close to the lowest
frequencies probed by the data and, as seen in Fig. \ref{doublecon12},
it is essentially unbounded down to the lowest measurable frequency at
the 68\% confidence level. These facts suggest that the second,
low-frequency, bend might not be required by the data and that the
improvement in the fit might be only due to the increased complexity
of the model fitted.

The likelihood of acceptance is better in the double-bend model than
in the single-bend model, 64 versus 44 \% respectively, but there are
more free parameters. In order to
determine the significance of this improvement, we performed
the following test.  Using the best-fitting single-bend PSD
parameters, we generated 300 realisations of the sets of \xte\ and \xmm\
lightcurves.  Each
realisation was then fitted with the best-fitting double-bend
parameters, exactly as was done with the real data, and the
distribution of their fit probabilities was constructed. We found that
121 out of the 300 single-bend simulations have a higher fit
probability than the real data, when fitted with the double-bend
model. Therefore, we conclude that the improvement in fit probability
is no more than may be
expected from fitting a model which is more complicated than required
by the data: the double-bend model does not represent a significant improvement.

\subsection{Single-bend power law with a Lorentzian component}

We finally consider whether the observed PSD might be best-described by
adding a Lorentzian component, such as are commonly used to describe
broad-band noise components in GBHs \citep[e.g.][]{Nowak:2000qf}, to
the single-bend power law.  We are motivated to consider this
possibility because the PSD of the intense-sampling \xte\ light curve
is not very well described by either the single- or double-bend power
law model.  Visual inspection of this light curve, shown in
Fig. \ref{intlc}, suggests that the variability is strongly
concentrated on time-scales of around a day, or equivalently,
frequencies around $10^{-5}$ Hz, which is confirmed by the peak seen
in the corresponding section of the PSD, and the drop in the same PSD
at lower frequencies ($\sim 10^{-6}$Hz). The long-term monitoring
PSDs, however, do not show a dip at $10^{-6}$Hz, creating a large
discrepancy in the PSD measurements at this frequency. A strongly
peaked component in the underlying PSD, at $\sim10^{-5}$ Hz, could
produce the observed features. Such a component would appear as a peak
in a PSD that covered frequencies above and below its peak-frequency,
but would be insufficiently sampled by the long-term monitoring
campaigns; thus, its power would be aliased into the highest
frequencies of the longer time-scale data, making them rise above the
underlying model level and causing the apparent disparity.

\begin{figure}
\psfig{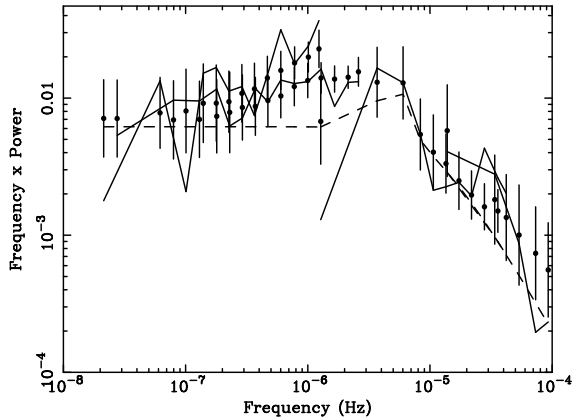}
\caption{The best-fitting single-bend power law with a Lorentzian
component. The fit was done using the
entire data set but here we only show the Lorentzian region. As
before, solid lines represent the real data PSD, dashed lines
represent the best-fitting model and markers with error bars represent
the model distorted by sampling effets. The Lorentzian feature in the
model can reproduce qualitatively the spurious power at the high
frequency end of the long-term monitoring data and the turn down
effect observed in the intensive-sampling data.}  \label{lorvfv}
\end{figure}

The Lorentzian profile is described by:
 
\begin{equation*}
P_{\rm Lor}(\nu)=\frac{A Q \nu_c}{\nu_c^2+4Q^2(\nu_c-\nu)^2},
\end{equation*}

\noindent
where the centroid frequency $\nu_c$ is related to the peak-frequency
$\nu_p$ by $\nu_p=\nu_c\sqrt{1+1/4Q^2}$ and the quality factor Q is
equal to $\nu_c$ divided by the full width at half maximum of the
Lorentzian. The variable $A$ parameterizes the relative contribution of
the power law and Lorentzian components to the total rms.
Fitting a Lorentzian component in addition to a single-bend power law
provides a good fit (P=52 \%).  The best-fitting Lorentzian contributes
20\% of the variance in the frequency range probed and its
best-fitting parameters are quoted in Table \ref{results_lor}.
Fig. \ref{lorvfv} shows the observed PSD compared with the
best-fitting single-bend power law model plus
a Lorentzian component. The
Lorentzian feature in the model can reproduce qualitatively the
spurious power at the high frequency end of the long-term monitoring
data and the turn down effect observed in the intensive-sampling data.

\begin{table*} \centerline{} \begin{tabular}{|c||c||c||c||c||c||c|}
\hline {\bf $\nu_p$} & {\bf $\nu_B$} & {\bf $Q$} & {\bf $A$} & {\bf
$\alpha_{L}$} & {\bf $\alpha_{H}$} & {\bf Acceptance}\\ {(Hz)}& {(Hz)}
& {} & {} & {} & {} & {(\%)}\\ \hline $4.8^{+*}_{-0.8}\times 10^{-6}$
& $1.1^{+0.6}_{-0.4}\times 10^{-5}$ & $5.1^{+*}_{-3.6}$ &
$0.9^{+*}_{-0.7}$ & $-1.0^{+*}_{-*}$ & $-2.6^{+*}_{-*}$& {52.3}\\
\hline \end{tabular} \caption{Best fit single-bend power law with
Lorentzian component model parameters to the combined \xte\ and \xmm\
PSD of NGC~3783, where $\nu_p$ is the Lorentzian peak frequency, $Q$
is its quality factor, $\nu_B$ is the power law bend frequency and
$\alpha_{L}$ and $\alpha_{H}$ are the power law slopes below and above
the bend, respectively.  The errors are calculated from the 68\%
confidence intervals, and an asterisk indicates that the limit is unconstrained.}  
\label{results_lor} \end{table*}

To determine the significance of the Lorentzian component fit we
repeated the procedure used in determining the significance of the
double-bend model. We found that 222 of the 300 single-bend simulated
PSDs have a higher fit probability than the data, when fitted with the
single-bend power law plus Lorentzian model.  This result indicates
that the increase in fit probability could be due to the added
complexity of the model, and that the improvement in the fit over a
simple bending power law is not significant.

\section{Discussion and Conclusions}

We have combined our own new \xte\ monitoring data with archival \xte\ and
\xmm\ observations to construct a high-quality PSD of NGC~3783
spanning five decades in frequency. 

We find that a `soft' state model, with a single bend at
$6.2\times10^{-6}$ Hz, similar to that found earlier by
\cite{Markowitz:2003gm}, a power law of slope approximately -$0.8$
extending over almost three decades in frequency below the bend, and
slope above the bend of approximately -2.6 is a good fit to the
data. We also find that a `hard' state model, with a double bend, fits
the data, as does a model with a single bend plus an additional
Lorentzian component. However the improvement in fit is marginal and,
given the additional free parameters, is not significant. Thus we
conclude that a simple `soft' state model provides the most likely
explanation of the data.

Assuming a mass of $3\times 10^{7} M_\odot$ for NGC~3783
\citep{Peterson:2004ve}, and an accretion rate of 7\% of the Eddington
limit (\citealt{Uttley:2005bh}, based on \citealt{Woo:2002cr}), then NGC~3783
is still in good agreement with the scaling of PSD break
timescale as $\sim M/\dot{m}_{E}$ between AGN and GBHs found by
\cite{McHardy:2006fk}.

Our new fits, show that the PSD of NGC~3783 is perfectly
consistent with a single-bend power law with low-frequency slope of
-1, in contrast with the earlier result of \cite{Markowitz:2003gm},
who found that a similar model was rejected tentatively at $\sim98$\% confidence.  The
difference can be understood in terms of the improved long-term
data. Our new \xte\ monitoring observations occur every 2 days,
compared to 4 days previously, thereby increasing the long term \xte\
data set by a factor 2.6 and, in particular, providing overlap at high
frequencies with the \xte\ intensive monitoring data. The drop in
long-timescale variability power, evident in the older long term
monitoring data is not reproduced by the new long-term monitoring
data, showing that this drop could be just a statistical fluctuation.
In addition, the very high frequencies are better constrained by the 2
orbits of
\xmm\ data than by the earlier \chandra\ data used by
\cite{Markowitz:2003gm}. 

The classification of the PSD as being `soft' state means that NGC~3783
is no longer considered unusual amongst AGN. The fact that this AGN is
radio-quiet strongly supports the analogy with GBHs
in the soft state. Also the accretion rate of
NGC~3783 (\me=0.07) (\citealt{Uttley:2005bh}, based on
\citealt{Woo:2002cr}) is similar to that of other AGN with soft-state PSDs
(e.g. NGC~3227 \citealt{Uttley:2005bh}, NGC~4051
\citealt{McHardy:2004hv}, MCG-6-30-15 \citealt{McHardy:2005pe}). 
This accretion rate is above the rate at which the persistent GBH 
Cyg~X-1 transits between hard and soft states in either direction  and
at which other GBHs transit from the soft to hard state (\me=0.02)
\citep{Maccarone:2003uq,Maccarone:2003kx}. We note that other transient GBHs in
outburst, where the variable power law emission in the soft state PSD is weak,
can remain in the hard state to much higher
accretion rates ($\sim$ 2--50\% \citealt{Homan:2005fk}) but it is not
clear whether we should expect similar PSD shapes to AGN for such
outbursting sources.
Thus NGC~3783 remains compatible with other
moderately accreting AGN in being analogous to Cyg~X-1 in the soft
state.  It is, of course, possible that the transition rate might not
be independent of mass.  Observations do not yet greatly constrain the
transition rate as a function of mass but the abscence of large
deviations from the so-called `fundamental' plane of radio luminosity,
X-ray luminosity and black hole mass
\citep{Merloni:2003kx,Falcke:2004ko} argues against a large spread in
the transistion accretion-rates (e.g. see \citealt{Kording:2006fk}).  In
the case of Seyfert galaxy NGC~3227, the accretion rate is $\sim$ 1--2\% and
a `soft' state PSD is measured \citep{Uttley:2005bh}, which suggests
that the transition accretion-rate in AGN should be at or below that
value.

Our observations, which show that NGC~3783 does not have a highly
unusual PSD, therefore confirm the growing similarities between
AGN and Galactic black hole systems and leave only Arakelian 564,
which is probably a very high state object, as the only AGN showing
clear double breaks (or multiple Lorentzians) in its PSD
(e.g. \citealt{Arevalo:2006dq}, M$^{\rm c}$Hardy et al. in prep.).

\section*{Acknowledgements}

We would like to thank the referee, Chris Done, for useful comments and suggestions. This research has made use of the data obtained from the High Energy Astrophysics Science Archive Research Center (HEASARC), provided by NASA's Goddard Space Flight Center.  We would like to thank Information Systems Services (ISS) at the University of Southampton for the use of their Beowulf cluster, \textit{Iridis2}.  PU acknowledges support from a Marie Curie Inter-European Research Fellowship.

\bibliographystyle{mn2e}
\bibliography{mn-jour,references}

\label{lastpage}
\end{document}